\documentclass[aps,prl,twocolumn,showpacs,superscriptaddress]{revtex4}
\usepackage{graphicx}
\begin{document}

%\preprint{}

\title{Field-induced Bose-Einstein Condensation of triplons up to 8 K in Sr$_3$Cr$_2$O$_8$}
     \author{A.~A.~Aczel}
     \affiliation{Department of Physics and Astronomy, McMaster University, Hamilton, Ontario, Canada, L8S 4M1}
     \author{Y.~Kohama}
     \affiliation{National High Magnetic Field Laboratory, Los Alamos National Laboratory, Los Alamos, New Mexico 87545, USA}
     \author{C.~Marcenat}
     \affiliation{CEA-Grenoble, Institut Nanosciences et Cryog$\acute{e}$nie, SPSMS-LATEQS, 17 rue des Martyrs, 38054 Grenoble Cedex 9, France}     
     \author{F.~Weickert}
     \affiliation{Max Planck Institute for Chemical Physics of Solids, Dresden, Germany} 
     \author{M.~Jaime}
     \affiliation{National High Magnetic Field Laboratory, Los Alamos National Laboratory, Los Alamos, New Mexico 87545, USA} 
     \author{O.~E.~Ayala-Valenzuela}
     \affiliation{National High Magnetic Field Laboratory, Los Alamos National Laboratory, Los Alamos, New Mexico 87545, USA}
     \author{R.~D.~McDonald} 
     \affiliation{National High Magnetic Field Laboratory, Los Alamos National Laboratory, Los Alamos, New Mexico 87545, USA}
     \author{S.~D.~Selesnic}
     \affiliation{Department of Physics and Astronomy, McMaster University, Hamilton, Ontario, Canada, L8S 4M1}
     \author{H.~A.~Dabkowska}
     \affiliation{Brockhouse Institute for Materials Research, McMaster University, Hamilton, Ontario, Canada, L8S 4M1} 
     \author{G.~M.~Luke}
     \affiliation{Department of Physics and Astronomy, McMaster University, Hamilton, Ontario, Canada, L8S 4M1}
     \affiliation{Brockhouse Institute for Materials Research, McMaster University, Hamilton, Ontario, Canada, L8S 4M1} 
     \affiliation{Canadian Institute of Advanced Research, Toronto, Ontario, Canada, M5G 1Z8}

\date{\today}

\begin{abstract}
Single crystals of the spin dimer system Sr$_3$Cr$_2$O$_8$ have been grown for the first time. Magnetization, heat capacity, and magnetocaloric effect 
data up to 65 T reveal magnetic order between applied fields of H$_{c1}$~$\sim$~30.4~T and H$_{c2}$~$\sim$~62~T. This field-induced order persists up to 
T$^{max}_c$~$\sim$~8~K at H~$\sim$~44~T, the highest observed in any quantum magnet where H$_{c2}$ is experimentally-accessible. We fit the 
temperature-field phase diagram boundary close to H$_{c1}$ using the expression T$_c=$~A(H-H$_{c1})^\nu$. The exponent $\nu=$~0.65(2), obtained at 
temperatures much smaller than T$^{max}_c$, is that of the 3D Bose-Einstein condensate (BEC) universality class. This finding strongly suggests that 
Sr$_3$Cr$_2$O$_8$ is a new realization of a triplon BEC where the universal regimes corresponding to both H$_{c1}$ and H$_{c2}$ are accessible at $^4$He 
temperatures.
\end{abstract}

%\keywords{}
\maketitle

%para 1

The Bose-Einstein condensate (BEC) is a unique state of matter in which a macroscopic number of bosons collapse into a global (identical) quantum 
state\cite{02_cornell, 02_ketterle}. Although superfluid $^4$He\cite{98_wyatt} is possibly the most studied BEC system, ultracold Rb gas proved to be the 
first true experimental realization of this exotic state of matter\cite{02_cornell}. Since the original discovery was made, other BECs have been identified 
in optical lattices, superconductors\cite{book_Pitaevskii}, and optically-pumped magnon gases\cite{06_demokritov} where the phenomenon of BEC can be 
studied at room temperature.
 
Giamarchi and Tsvelik first showed that the Hamiltonians of quantum antiferromagnets and BECs were directly related 
by a mapping transformation\cite{99_giamarchi}, following up on earlier work by Matsubara and Matsuda\cite{56_matsubara}. Although real lattices always 
have imperfections and symmetry-breaking terms, in some materials these can be neglected since they only affect their properties 
below the experimentally-accessible window. This then suggests that a very good approximation to a BEC can be achieved in solid-state magnetic systems.
In particular, spin dimer systems were perceived to be good candidates 
to display BEC behaviour. These systems exhibit spin-singlet ground states with a spin gap to the first excited triplet state. This 
gap can be closed by applying a large enough magnetic field, resulting in the generation of a macroscopic number of 
triplet excitations (triplons). If the kinetic energy terms dominate
the potential energy terms in the U(1) invariant Hamiltonian of such a system, the resulting ground state can be described as a BEC. 

TlCuCl$_3$ was the first spin dimer system to show evidence for a triplon BEC ground state\cite{99_oosawa, 00_nikuni}, and this has stimulated a flurry of 
research in triplon BECs\cite{04_jaime, 06_zapf, 09_aczel, 09_kofu, 06_sebastian}. Currently, the primary goal is to search for systems displaying
BEC behaviour in the space of externally-controlled parameters completely accessible in the laboratory. This allows the associated quantum phase 
transitions to be studied experimentally. 

The field-induced order in a triplon BEC is nothing but the well-known XY-AFM (easy-plane antiferromagnetic) state. Such a system can either be visualized 
as a lattice gradually filling with hard-core triplons, or as an XY-AFM where the in-plane component of the canted spins varies from zero 
in the paramagnetic state, to a maximum at approximately midway between H$_{c1}$ and H$_{c2}$, to zero again in the fully-polarized state. The external 
control parameter in both cases is the applied magnetic field, and the order parameter is the number of triplons or, 
equivalently, the magnitude of the staggered moment M$_{xy}$. The (H,T) phase boundary in ideal 3D systems follows the expression 
T$_c=$~A(H-H$_{c})^{2/3}$ close to the critical fields H$_{c1}$ and H$_{c2}$. Symmetry breaking terms can affect real systems by changing the universality 
class from XY-AFM (easy plane) to Ising (easy axis), in which case the critical exponent changes from 2/3 to 1/2, or by changing the dimensionality of 
the system\cite{08_giamarchi}.   
 
In this letter, we present the first evidence for a BEC ground state in the spin dimer system Sr$_3$Cr$_2$O$_8$ through a combination of high field 
magnetization (M(H)), heat capacity, and magnetocaloric effect (MCE) measurements. This is one of a series of isostructural
spin dimer compounds with the general formula A$_3$M$_2$O$_8$\cite{06_nakajima, 01_uchida, 07_singh} where A~$=$~Ba or 
Sr and M~$=$~Cr or Mn. These compounds crystallize in the R-3m space group at room temperature, and the magnetic M$^{5+}$ ions are tetrahedrally- 
coordinated with O$^{2-}$ ions. The M$^{5+}$ ions may carry spins of either S~$=$~1/2 or 1, and as a result of having only one nearest neighbour 
(NN) magnetic ion they form spin dimers along the c-axis. While the A$^{2+}$ ions are essentially 
isolated from the MO$^{4-}$ tetrahedra, they seem to play an important role in controlling the NN M-M distance and the value of the associated spin gap. 
For example, while Ba$_3$Cr$_2$O$_8$ has a NN Cr-Cr distance of 3.9765 \AA~and a spin gap of 
15.6(3)~K\cite{08_aczel}, Sr$_3$Cr$_2$O$_8$ has a smaller NN Cr-Cr distance of 3.842 \AA~and hence a larger spin gap of 
61.9(1)~K\cite{07_singh}. This is consistent with the expectation that a smaller NN distance between the magnetic ions should yield a larger 
antiferromagnetic intradimer exchange interaction. The stronger intradimer exchange interaction in the Sr-Cr system pushes the temperature scale of 
the magnetic order to the highest values observed yet in a system where H$_{c2}$ is experimentally-accessible. 

Single crystals of Sr$_3$Cr$_2$O$_8$ were grown by the optical floating zone method at McMaster University. We first  
mixed together stoichiometric quantities of SrCO$_3$ and Cr$_2$O$_3$. The resulting powder was shaped into 10 - 12 cm long rods, which were 
annealed for 72 hours in air at 1200$^{\circ}$C and then quenched to room temperature. The rods cracked during the annealing process, so they were 
remade following this and subsequently annealed for $\sim$5 hours in high purity Ar gas at 1200$^{\circ}$C to strengthen them for the  
floating zone growths. Since Sr$_3$Cr$_2$O$_8$ is congruently melting in Ar gas, the crystals were grown at a very high rate of 20 mm/h with an Ar 
pressure of $\sim$170~kPa and a temperature of 1470$^{\circ}$C. The phase purity of the resulting crystals was confirmed by 
powder x-ray diffraction, and they were aligned by Laue x-ray diffraction. 

The anisotropy of the g-factor was measured via electron paramagnetic resonance (EPR) at 70 GHz using a resonant cavity mounted in a cryogenic goniometer, 
so as to rotate the sample and resonant environment in situ. The temperature was maintained at 30~K to ensure sufficient thermal population of the triplet
state using a $^4$He flow cryostat. The magnetic field was applied using a superconducting solenoid and calibrated using the conventional EPR-marker DPPH. 
The g-factors were found to be g$_c$~$=$~1.938(6) and g$_{ab}$~$=$1.950(1). 
 
\begin{figure}[t]
\includegraphics[width=3.5in,angle=0]{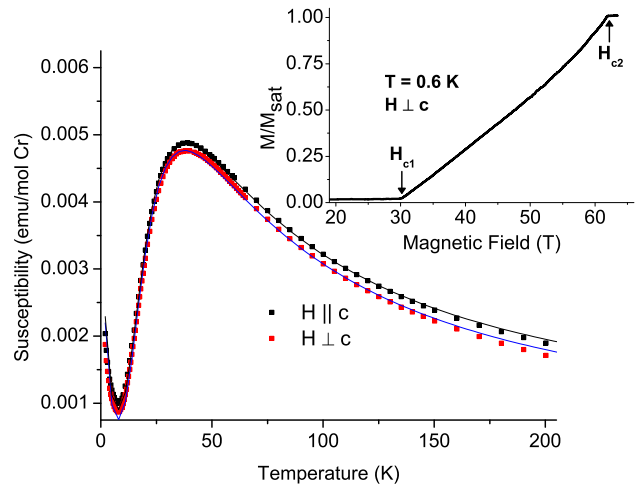}
\caption{\label{fig1}(color online)
dc susceptibility measurements of Sr$_3$Cr$_2$O$_8$ with an applied field of 1~T. The data was fit to an isolated dimer model. The inset shows the 
magnetization M(H) as measured in a 65~T short pulse magnet.}
\end{figure}

Fig.~\ref{fig1} depicts dc susceptibility measurements of our crystals. At high temperatures, the data follows a Curie-Weiss law. Below $\sim$~40~K, the 
susceptibility decreases sharply with decreasing temperature, which is characteristic of non-magnetic spin-singlet ground state systems. The susceptibility
measurements also reveal that this system is nearly isotropic. We fit the susceptibility data to an interacting dimer model of the form:
\begin{equation}
\chi_M=\frac{N_A(\mu_Bg)^2}{k_bT(3+\exp(J_0/T)+J'/T)}+\chi_0+\frac{A}{T}
\end{equation}
where N$_A$ is Avogadro's number, $\mu_B$ is the Bohr magneton, J$_0$ is the intradimer exchange constant, and J' is the sum of the interdimer exchange 
constants. The last two terms represent contributions from Van Vleck paramagnetism/core diamagnetism and impurity/defect spins respectively. We fixed the
g-factors in the fits to the values obtained by EPR. From the fitting, we found only $\sim$0.8\% free S~$=$~1/2 moments, and estimated the exchange 
constants to be J~$=$~61.5(3)~K and J'~$=$12(2)~K. Note that the latter are actually best determined by inelastic neutron scattering\cite{09_lake}. 

Magnetization data was collected using an extraction magnetometer in conjunction with a $^3$He fridge and a 65 T short pulse (35 ms) magnet at 
the National High Magnetic Field Laboratory (NHMFL) in Los Alamos, New Mexico. The Fig. 1 inset shows M(H) for the H$\perp$$\hat{c}$ orientation while 
sweeping the field up. M(H) has a very small value up to H$_{c1}=$~30.4~T 
due to defects/impurities in the sample. Above H$_{c1}=$~30.4 T, 
the spin gap is closed resulting in a magnetically-ordered regime between H$_{c1}$ and H$_{c2}=$~62~T where M(H) 
increases roughly linearly with H. Finally, above H$_{c2}$ M(H) saturates as the system reaches the fully-polarized state. 

To further characterize the field-induced order, heat capacity and MCE measurements were performed 
using a home-built calorimeter\cite{08_samulon, 09_aczel}. These measurements were 
conducted in a $^3$He fridge placed in the core of a 35~T resistive magnet at the NHMFL in Tallahassee, Florida. The MCE data was collected 
for the H$\perp$$\hat{c}$ orientation, sweeping the field both up and down at 3~T/min. The quasi-adiabatic MCE yields a sharp change in the sample 
temperature when entering or leaving a magnetically-ordered state to ensure entropy conservation\cite{06_silhanek}. 

Fig. 2(a) shows a color contour plot of the entropy as a function of temperature and field. $\Delta$S was calculated from the MCE data using the equation:
\begin{equation}
\Delta S=\int\frac{\kappa(T-T_{bath})}{T}dt
\end{equation}
where $\kappa$ is the thermal conductivity of the thermal link in the calorimeter, $T$ is the sample temperature, and $T_{bath}$ is the bath temperature. 
We measured $\kappa$ as a function of $T$ at both 0 and 33~T and found very little difference in the two curves, so the 33~T thermal conductivity data 
was used in all our calculations. The absolute entropy was then obtained by a calibration process using 31.5~T heat capacity data. We chose to
plot S/T to improve the contrast since the entropy associated with the transition drops quickly with decreasing temperature. As 
one crosses the phase boundary at a constant temperature, a peak in the entropy is clearly evident. This results because both the low-field gapped phase 
and the ordered phase above the transition should have small entropies. However, in the proximity of the transition we likely have  
fluctuations of the order parameter characteristic of a second order transition and hence a much higher entropy. 

\begin{figure}[t]
\includegraphics[width=3.2in,angle=0]{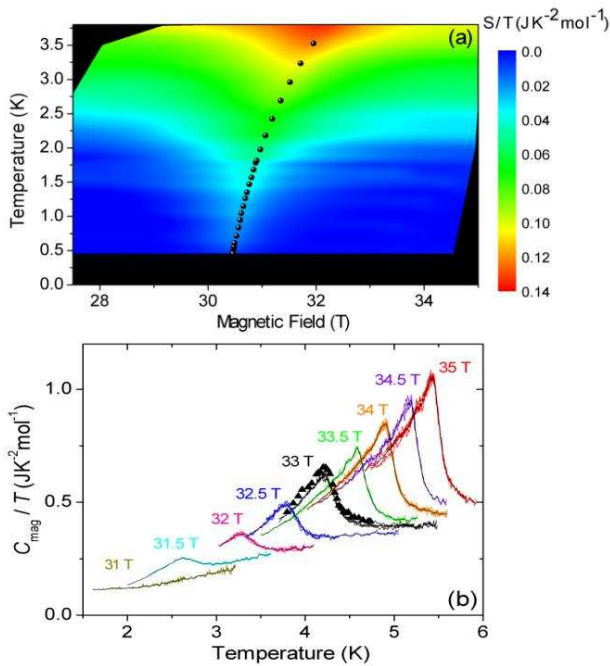}
\caption{\label{fig2}(color online)
(a) Color contour plot of the entropy S/T of Sr$_3$Cr$_2$O$_8$, as calculated from the MCE data. The phase 
boundary determined by MCE is shown with black dots. (b) Selected heat capacity data as obtained by 
the dual slope method. For 33 T, data was also collected by the standard relaxation method (black triangles).}
\end{figure}

Heat capacity measurements are presented in Fig. 2(b) for the H$\perp$$\hat{c}$ orientation. The majority of the 
data was collected using the dual slope method with a large delta T\cite{86_riegel}. The main advantage of this technique is that data 
can be collected quickly which is important for efficient use of power in high field resistive magnets. However, this often comes at a cost 
of reduced accuracy, and so data is generally collected for at least one applied field using the standard relaxation technique to compare the two methods 
directly. In the present case, this was done at 33~T (black triangles in Fig. 2(b)). The excellent agreement between 
the techniques suggests that the dual slope method estimates the heat capacity well in Sr$_3$Cr$_2$O$_8$. 

The emergence of a lambda-like anomaly was clearly observed with an increasing applied field beyond H$_{c1}\sim$~30.4~T; this provides unambiguous evidence
for the field-induced order in this system. The lambda anomaly is small for fields close to H$_{c1}$, but becomes much more prominent with 
increasing applied field as seen in Fig. 2(b). Heat capacity data was also collected for the H$\parallel$$\hat{c}$ orientation. Similar lambda anomalies 
were observed in that case, and the transition temperatures were systematically shifted down by only $\sim$ 0.2~K for a given applied field. This indicates
that Sr$_3$Cr$_2$O$_8$ is a very isotropic, as previously suggested by the susceptibility measurements.

The spin gap in Sr$_3$Cr$_2$O$_8$ is quite large, and so DC fields provided by resistive magnets can access only a very small region of the phase diagram. 
For this reason, we also performed MCE measurements in both the 50 T mid-pulse (300 ms) and 60 T long-pulse (2 s) 
magnets in Los Alamos. For these measurements, a tiny Sr$_3$Cr$_2$O$_8$ single 
crystal (200 x 100 x 50 $\mu m^3$) was inserted into a small, home-built calorimeter\cite{09_kohama}. 
To the best of our knowledge, these were the first successful MCE experiments performed in a pulsed field facility. 

Fig. 3 presents a phase diagram for the H$\perp\hat{c}$ orientation of Sr$_3$Cr$_2$O$_8$ as obtained from our specific heat, MCE and M(H) experiments.  
The asymmetric nature of the dome is quite unusual for a triplon BEC system; further studies are required to understand this feature. 
From the heat capacity measurements, the ordering temperature for a given field was taken as the peak of the lambda-like feature.
From the MCE data, the ordering temperature for a given field was determined by locating the extrema in the first derivative of the sample 
temperature with respect to field\cite{08_samulon, 09_aczel}. Some selected upsweep MCE curves are superimposed on the phase diagram. For clarity 
reasons we show only one downsweep MCE curve; the sample temperature changes in the opposite manner to the upsweep data near 
the transition as expected. No irreversible/dissipative mechanisms typical for a first order phase transition were observed. Many MCE scans were completed 
at low temperatures so it would be possible to determine a critical exponent. The fitted data all must lie in the universal regime for the 
obtained critical exponent to be meaningful, and an earlier Monte Carlo simulation study\cite{04_kawashima} suggested that the universal regime extends no
higher than $\sim$0.4T$_c^{max}$, where T$_c^{max}$ is the maximum temperature of the magnetically-ordered regime. In the present case, we have determined 
that the top of the dome lies around 8 K. To ensure that we stayed within the universal regime, only the MCE data below 2.7~K was fit to a power 
law of the form: T$_c=$~A(H-H$_{c1})^\nu$. 

\begin{figure}[t]
\includegraphics[width=3.4in,angle=0]{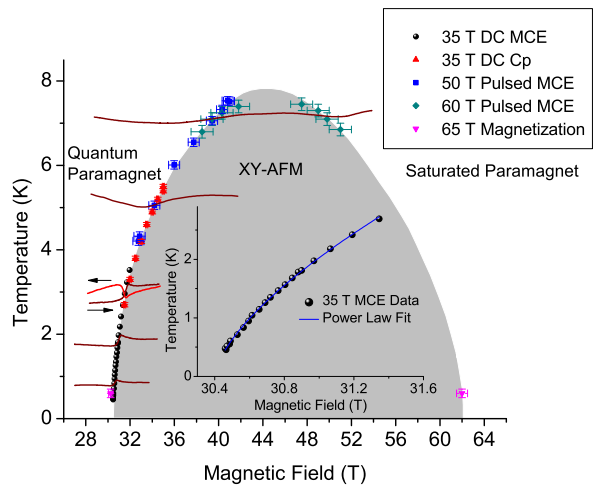}
\caption{\label{fig3}(color online)
Sr$_3$Cr$_2$O$_8$ phase diagram with selected MCE traces superimposed and the ordered region shaded as a guide to the eye. Due to small temperature
changes at the transitions, the MCE traces have 
been magnified. The inset shows the result of fitting the data near H$_{c1}$ to a power law.}
\end{figure}

Although one can perform a three parameter fit, H$_{c1}$ and $\nu$ are not independent variables. For this reason, we resorted to a series of two 
parameter fits where H$_{c1}$ was systematically fixed at different values and $\nu$ was determined from the fitting. Our best fit yielded $\nu=$0.65(2) 
with a corresponding H$_{c1}$~$=$~30.40(1)~T. The former is in agreement with the expected value of 2/3 for a 3D BEC universality class, strongly 
suggesting that Sr$_3$Cr$_2$O$_8$ is a realization of a new triplon BEC.  

Finally, let us comment here on the nature of the magnetically-ordered state in Sr$_3$Cr$_2$O$_8$. The spin structures in the
ordered states of the isostructural systems Ba$_3$Cr$_2$O$_8$\cite{09_kofu} and Ba$_3$Mn$_2$O$_8$\cite{08_samulon} have both been determined previously, 
and the results were quite different. In the case of Ba$_3$Mn$_2$O$_8$, a combination of single ion anisotropy and geometric frustration lead 
to both an incommensurate spiral spin phase and an Ising-type modulated structure in the ordered regime\cite{08_samulon}. The shape and structure of the 
phase diagram for that material depends on the orientation of the system relative to the applied field. In sharp contrast, neutron measurements revealed a 
commensurate, collinear spin structure in Ba$_3$Cr$_2$O$_8$\cite{09_kofu}. This is likely a result of the geometric frustration being relieved due to a 
structural distortion courtesy of a co-operative Jahn-Teller effect\cite{08_kofu}. A similar Jahn-Teller effect has been reported for
Sr$_3$Cr$_2$O$_8$\cite{08_chapon}, suggesting that the spin structure of the ordered state is very similar to the case of Ba$_3$Cr$_2$O$_8$. 

In summary, we have determined the full (H,T) phase diagram for Sr$_3$Cr$_2$O$_8$ from both thermodynamic measurements and high-field magnetization. This 
phase diagram provides direct evidence for field-induced order in this system between H$_{c1}\sim$~30.4~T and H$_{c2}$~$\sim$~62~T, with an unprecedented 
maximum ordering temperature T$_c^{max}$~$\sim$~8~K. The phase boundary near H$_{c1}$ was fit to a power law. Our obtained critical exponent of 0.65(2) 
agrees with the 3D BEC universality class, and so this strongly suggests that 
Sr$_3$Cr$_2$O$_8$ is a realization of a new triplon BEC system. Since a large portion of the universal regime in this material can be accessed via $^3$He 
systems, both the H$_{c1}$ and H$_{c2}$ quantum critical regimes can be studied in detail. This may help us to understand the asymmetric nature of the 
phase diagram and other more general aspects of triplon BEC phenomena. Note that no experimental criticality studies near 
H$_{c2}$ have been reported for triplon BEC systems to-date.

\begin{acknowledgments}
We acknowledge useful discussions with C.D. Batista and technical/experimental assistance from A.B. Dabkowski, F. Balakirev, and N. Harrison. Research at 
McMaster University is supported by NSERC and CIFAR. Research at NHMFL is supported by the National Science Foundation, the Department of Energy, and the 
State of Florida. 
\end{acknowledgments}

\vfill \eject
\end{document}